\documentclass[iop,onecolumn,onecolappendix,numberedappendix]{emulateapj}

\usepackage{amsmath,amsthm}
\usepackage{booktabs}
\usepackage{threeparttable}
\usepackage{tabularx}
\usepackage{xcolor}

\usepackage{times}
\usepackage{graphicx}
\usepackage{comment}

\newcommand{\cs}{c_{\rm s}}
\newcommand{\Sigmag}{\Sigma_{\rm gas}}
\newcommand{\rhog}{\rho_{\rm gas}}
\newcommand{\Mg}{M_{\rm gas}}
\newcommand{\Mgt}{\overline{M}_{\rm gas}}
\newcommand{\Rg}{R_{\rm gas}}
\newcommand{\Rgt}{\overline{R}_{\rm gas}}
\newcommand{\Rgh}{R_{\rm gas/2}}
\newcommand{\Rght}{\overline{R}_{\rm gas/2}}

\newcommand{\taudyn}{\tau_{\rm dyn}}
\newcommand{\taurad}{\tau_{\rm cool}}
\newcommand{\tauSFC}{\tau_{\rm SF,C}}
\newcommand{\taujeans}{\tau_{\rm Jeans}}
\newcommand{\taurot}{\tau_{\rm rot}}
\newcommand{\drhost}{\dot\rho_*}
\newcommand{\drhostT}{\dot\rho_{\rm *,Q}}
\newcommand{\drhostC}{\dot\rho_{\rm *,C}}
\newcommand{\etaSFQ}{\eta_{\rm SF,Q}}
\newcommand{\etaSFC}{\eta_{\rm SF,C}}

\newcommand{\dSigtot}{\dot\Sigma_*}
\newcommand{\dSigQ}{\dot\Sigma_{\rm *,Q}}
\newcommand{\dSigC}{\dot\Sigma_{\rm *,C}}

\newcommand{\Mstar}{M_*}
\newcommand{\Rec}{<R_{\rm e}>}

\newcommand{\Mdstar}{M_{\rm d*}}
\newcommand{\Rdstar}{R_{\rm d*/2}}
\newcommand{\Rdmsh}{R_{\dot M_*/2}}
\newcommand{\Rdmsht}{\overline{R}_{\dot M_*/2}}

\newcommand{\ke}{k_{\rm e}}

\newcommand{\Ampl}{{\cal A}_*}

\newcommand{\Msun}{M_{\odot}}
\newcommand{\Msol}{M_{\odot}}
\newcommand{\kpc}{{\rm kpc}}
\newcommand{\pc}{{\rm pc}}
\newcommand{\yr}{{\rm yr}}

\graphicspath{{./}{figures/}}

\begin{document}

\title{Star formation inefficiency and Kennicutt-Schmidt laws in Early-Type Galaxies}

\author{Brian Jiang}
\affiliation{Department of Physics, Columbia University, 550 West 120th St, New York, NY 10027, USA}

\author{Luca Ciotti}
\affiliation{Department of Physics and Astronomy, University of Bologna, Bologna, Italy
}

\author{Zhaoming Gan}
\affiliation{New Mexico Consortium, Los Alamos, NM 87544, USA}
\affiliation{Department of Astronomy, Columbia University, 550 West 120th St, New York, NY 10027, USA}

\author{Jeremiah P. Ostriker}
\affiliation{Department of Astronomy, Columbia University, 550 West 120th St, New York, NY 10027, USA}
\affiliation{Department of Astrophysical Sciences, Princeton University, Princeton, NJ 08544, USA.}

\begin{abstract}

  Star formation in disk galaxies is observed to follow the empirical
  Kennicutt-Schmidt law, a power-law relationship between the surface
  density of gas $\Sigmag$ and the star formation rate per unit
  surface $\dSigtot$. In contrast to disk galaxies, early-type
  galaxies (ETGs) are typically associated with little to no star
  formation, and thus are usually termed ``quiescent''.  Recent
  observations, however, have noted the presence of massive gaseous
  cold disks in ETGs, raising the question as to why the conversion of
  gas into stars is truly inefficient. With the aid of the latest
  simulations, performed with our high-resolution hydrodynamic
  numerical code \texttt{MACER}, we reevaluate the traditional
  classification of ETGs as quiescent, dead galaxies. In fact, in
  the presence of moderate galaxy rotational support, circumnuclear
  gaseous disks of kpc size form following cooling episodes of the
  ISM, and are not destroyed by AGN feedback.  The issue of star
  formation in such disks is therefore unavoidable. In MACER we follow
  star formation by considering two channels, that of 1) Toomre instability and 2) gas
  cooling/Jeans instability.  We find that the resulting
  Kennicutt-Schmidt laws for the simulated ETGs reproduce the observed
  slope in disk galaxies, though with
  considerable scatter and lower normalization by a factor
  of $\approx 2$ or more for the highest mass galaxies. The Toomre
  instability is the main driver of the slope, while cooling/Jeans
  induced star formation dominates the central regions. Observational checks of our star formation
  predictions are thus essential for confirming the form of local star
  formation laws and reassessing star formation inefficiency in
  ETGs. The process we describe is similar to that of the star forming
  disks - of much lower luminosity - found in the central spheroidal
  regions of local spirals, such as the MW and M31.

\end{abstract}


\section{Introduction} 

Understanding how the physical properties of interstellar gas affect
star formation is important for developing models of galactic
evolution and possibly explaining the differences in star formation
rate (SFR) across different galaxy types. A robust empirical
correlation between the SFR and gas density in disk galaxies was first
reported by \cite{Schmidt_1959}, suggesting a (volumetric) star
formation law for the Milky Way well described by a power-law
$\drhost\propto \rhog^n$, with $2<n<3$.  Starting with Kennicutt's
compilation of H$\alpha$ measurements to trace star formation and HI
and CO data to trace atomic and molecular gas, it has been found that
both {\it global} (Kennicutt 1998) and {\it resolved} (
Kennicutt 1989, hereafter K89) laws relating SFR and gas surface
densities read $\dSigtot\propto \Sigmag^n$, with $1<n<3$, and
$n \approx 1.4$ as the most accepted value. This empirical law,
universally known as the Kennicutt-Schmidt law (hereafter KS) forms
the critical basis for current theoretical and numerical work on
common disk galaxies. Here and in the following, the {\it global} observables entail averaging the gas surface density and the star formation rate within some prescribed radius in the galactic disk, while the
{\it resolved} values of $\Sigmag$ and $\dSigtot$ are just
angular averages over annuli of prescribed inner and outer radii $R$
and $R+\Delta R$.

Due to the importance of the subject, it is not surprising that a huge
amount of observational, numerical, and theoretical work has been done
to elucidate the physics behind the observed star formation rates. As
the real phenomenon involves turbulent hydrodynamics, gravitational
instabilites, radiative transport, magnetic fields, etc., in general a
phenomenological approach is used (with considerable success) in the
attempt to capture the basic physical principles driving star
formation. In this framework, two main channels of star formation at work in relatively isolated galaxies (i.e., galaxies
where gas rich merging events and other environmental effects such as
near encounters, tidal stripping and harassment, can be neglected)
can be easily identified.  The first is related to local/global
instabilities in rotating gaseous disks (hereafter the ``Toomre instability'', see e.g. Binney \& Tremaine
2008). The second is the also well known criterion based on the
comparison between the cooling and the Jeans instability times
(hereafter the ``cooling/Jeans'' channel).  Both channels are almost
certainly at work in star formation episodes hosted by rotating,
massive cold gaseous disks.  Regarding the Toomre instability channel,
theory and simulations have suggested that in disk galaxies,
gravitational instability criterions \citep{Boissier_2003,
  Kennicutt_Jr__1998} accurately replicate both the KS power-law 
(see Leroy et al. 2008 for qualification) and
the cutoff threshold, while different disk thickness
\citep{Bacchini_2019}, turbulence \citep{Shetty_2008}, and shear
\citep{Davis_2014} can contribute to the observed scatter in
$n$. Additionally, efforts have also gone into examining modified KS
relations with terms including orbital velocities and velocity
dispersions of stars \citep{Schaye_2002}, with the aim of
finding a universal relationship between gas and SFR densities in all
types of star-forming galaxies (see Sun et al. 2023).  Regarding the
cooling/Jeans channel, it is obvious that in case of dense and cold
gas, the consequences of the associated short time scales are
inescapable, resulting in fast cooling, fragmentation, collapse of the ISM,
and consequent star formation. For example, numerical simulations
restricted to the cooling/Jeans channel (with no AGN feedback, and in
absence of Toomre instability) have shown that the resulting star
formation rates are in nice qualitative agreement with the observed
slope of the KS empirical law \cite{Negri_2014}.

Therefore, a {\it first} natural question is if the Toomre and the
cooling/Jeans channels are cooperative, or complementary, in the
process of star formation in rotating gaseous disks.  A {\it second},
more observationally motivated question, concerns the systematic
differences of star formation as a function of the morphological type
of the host galaxy. In fact, in contrast to disk galaxies, ETGs are
typically classified as ``red and dead" stellar systems, with quenched
star formation as a result of a post star-burst depletion of the
molecular gas reservoir \citep{Cappellari_2011,Baron_2022}, and
successive maintenance of galactic winds due to the cooperative
effects of SNIa, and recurrent AGN feedback events fueled by periodic
cooling of the ISM produced by the galaxy aging stellar population
(e.g., Ciotti et. al 1991 and Ciotti \& Ostriker 1997; see
  also Kim \& Pellegrini 2012).  However, in sufficiently massive
ETGs, the SNIa and AGN feedback effects are not strong enough to
maintain the galaxy devoid of the gas produced by the stars, and so in
such systems some central star formation will occur. Our work (Ciotti et al. 2022, hereafter C22) and that of others leads to the
conclusion that for massive ETGs ($\Mstar\approx 10^9$), roughly
$10^{11}\Msun$ of gas is ejected by feedback processes (SNIa and AGN
feedback) and a remaining $10^9\Msun$ falls to a central kpc disk in
what was initially termed ``cooling flow'' (e.g., see Fabian 1994,
Matthews \& Brighenti 2003)
and references therein. There it can fall to the center to feed the
AGN of be transformed locally into a (top heavy) population of new
stars, the most massive of which will eject, by dynamical processes,
some fraction of the gas from which they were born. In fact
observations show that ETGs possess some reservoir of cold gas, with
approximately 50 percent of massive ETGS (stellar mass
$\Mstar\gtrsim 10^{10} \;\Msol$) containing $10^6-10^9 \; \Msol$ of
cold gas in the form of atomic and molecular hydrogen (e.g., see
Cappellari et al. 2011, Li et al. 2020): while the gas reservoirs for
massive disk galaxies can reach $\approx 10^{10} - 10^{11} \Msol$
(Zhou et al. 2021, Saintonge \& Catinella 2022), validating the
picture of overall quiescent ETGs, non-negligible star formation seems
inevitable given the two channels above. Quite remarkably, it has been
noted that ETGs - for given amount of cold gas available - lie on a
significantly lower regime compared to starburst and disk galaxies on
the global KS relations \citep{Davis_2014}, implying that conversion
from gas to stars in ETGs is more inefficient per unit mass. This
inefficiency of star formation, despite non-negligible reservoirs of
cold gas, has been identified as one of the most persistent problems
in the field of star formation \citep{Peng_2015}.

As numerical simulations have demonstrated (Negri et al. 2015, C22),
ordered rotation enhances ISM instabilities and radiative cooling;
because the cooling rate grows quadratically with density, the
increasing ISM density triggers (as noted above) a cooling flow. As
the catastrophically cooling ISM looses its thermal pressure, it
accumulates onto a circumnuclear disk due to the angular momentum
barrier. Though the gaseous disk masses of ETGs are typically one to
two orders of magnitude less than what has been observed for spiral
galaxies, disk instabilities and density-dependent cooling rates
dictate that nonzero star formation is inevitable. ETGs, however, are
noted to be significantly less efficient in turning this cold gas to
stars in comparison to spiral galaxies. The reason why star formation is so inefficient in
ETGs, however, is not well understood; possible explanations involve
the stabilizing influence of stellar population on gaseous disks
\citep{2007ApJ...669..232K}, low disk self-gravity and increased shear
\citep{2013MNRAS.432.1914M}, and counter-rotation of stellar and
gaseous populations \citep{2017MNRAS.471L..87O}. Confirmations of
theoretical predictions via observations of star formation in ETGs are
hampered by the lack of resolution of inner star-forming disks; recent
far-infrared observations by \cite{Baron_2022} have suggested that a
potential resolution of this problem can be that much of the star
formation is obscured.

In this paper we address the two problems mentioned above, tacking
advantage of the latest suite of simulations produced with the current
version of our high-resolution hydrodynamical simulation code
\texttt{MACER}, which includes numerical algorithms for the radiative
cooling of the ISM, the formation of cold gaseous disks in the
presence of ordered rotation, and star formation following simple
feedback of various forms and robust input physics. We notice that the
resolution of \texttt{MACER} (parsec scale in the inner regions)
greatly exceeds that of most observations and typical cosmological
simulations, enabling the simulation and analysis of star forming
disks with half mass radius close to the galactic nucleus.

We are therefore in a good position to study the star formation
inefficiency problem in ETGs and the form of local star formation
laws. Applying \texttt{MACER} towards analyzing the formation of dense
star-forming disks close to the galactic nucleus, we have concluded
that the observed inefficiency in SFR can be partially explained due
to an underestimation of total star formation and the rapid ejection
of cold gas. While the degree of star formation is certainly less than
that of disk galaxies, we note that the formation of cold,
centrifugally supported equatorial disks due to the conservation of
angular momentum in the ISM of rotating ETGs creates an environment
conductive for star formation following cooling flows (see C22). Dense
stellar disks embedded in these cold gaseous disks, if they exist,
would lie close to the galactic center beyond the resolution of what
most resolved observations have attempted, and would be vulnerable to
destruction via late-time (dry) merging. Gas lost from evolving stars
is considerable and can be estimated at $\approx 10\%$ of the mass of
the remaining stars (Kim \& Pellegrini 2012).

This paper is organized as follows: in Section 2, we provide a brief
summary of the relevant numerical physics of \texttt{MACER}; in
Section 3.1, we discuss the position of ETGs on the global KS law with
respect to the cooling flow problem; and in Section 3.2-3, we discuss
resolved KS law to examine the dynamics of local SF, followed by a
summary of results.

\section{The numerical code and input physics}

In this paper we address some aspect of star formation in ETGs by a
detailed analysis of the suite of numerical simulations presented in
C22, obtained with the latest 2D version of our high-resolution code
\texttt{MACER} (Massive AGN Controlled Ellipticals Resolved), built
upon \texttt{Athena++} (version 1.0.0, see Stone et al. 2020).  With
this code we solve the Eulerian hydrodynamical equations for the ISM of
ETGs with a central supermassive black hole, while including the effects of circumgalactic (CGM) gaseous infall, dust grain evolution,
stellar feedback, AGN activity, and the metallic evolution of the ISM and also providing an accurate treatment of stellar dynamics. At this
stage, simulations are performed under the assumption of axisymmetry,
and phenomena such as merging, and tidal interactions with other galaxies
are not taken into account. For a detailed description of the input
physics, and other technical aspects, see C22 and references therein.

For the models discussed here, the hydrodynamical equations are solved
in spherical $(r,\vartheta)$ coordinates, while allowing for rotation
in the $\varphi$ direction. The outer boundary of the computational
domain is fixed to $r=250$ kpc from the galactic center on a
logarithmic radial grid, while the inner boundary is set to either 2.5
pc or 25 pc, the latter of which is used for a (relatively) fast
parameter-space exploration; we notice that the obtained resolution,
also in the less resolved case, is significantly finer than the
$\approx 150$ pc inner boundary of typical cosmological simulations and resolves the fiducial Bondi radius, allowing infall to the
central BH to be appropriately computed.

The galaxy models used are the axisymmetric JJe dynamical models
consisting of a moderately flat Jaffe ellipsoidal stellar distribution
(corresponding to an E3 galaxy) embedded in a dark matter halo,
resulting in a spherical Jaffe density profile (Ciotti et al. 2021,
C22). The additional gravitational field produced by a
quasi-isothermal DM halo is also considered, to model the effects of a
group/cluster halo.

We analyze the star formation properties in three families of models,
defined by their initial stellar mass, and fully described in C22 (see
Table 1 therein). In the family of {\it high mass} galaxy models (HM)
the initial stellar mass is $\Mstar = 7.8 \times 10^{11} \; \Msol$,
the edge-on circularized effective radius is $\Rec=11.8$ kpc, and the
stellar central velocity dispersion $\sigma_{*0}=312$ km s$^{-1}$.  In
the second family of {\it medium mass} galaxy models (MM) the initial
stellar mass is $\Mstar = 3.35 \times 10^{11} \; \Msol$, the edge-on
circularized effective radius is $\Rec=7.04$ kpc, and the stellar
central velocity dispersion $\sigma_{*0}=265$ km s$^{-1}$. In the
third family of {\it low mass} galaxy models (LM) the initial stellar
mass is $\Mstar = 1.54 \times 10^{11} \; \Msol$, the edge-on
circularized effective radius is $\Rec=4.57$ kpc, and the stellar
central velocity dispersion $\sigma_{*0}=223$ km s$^{-1}$.  All the
models are simulated over a time interval of $\Delta t = 12$ Gyr from
$t_0 = 2$ Gyr, i.e., we start our simulations after the galaxy
formation epoch.

In the presence of some galaxy rotation, the angular momentum
associated with the gas injected into the ISM by the evolving stars
(if not ejected by the galaxy as a galactic wind) leads to the
unavoidable formation of a cold and rotationally supported gaseous
disk in the galaxy equatorial plane, where star formation occurs (in
the simulations, the gravity of the resulting stellar disk is also
considered): of course, higher the rotation, bigger the gaseous disk
produced by ISM cooling, with important consequences for star
formation.  It is therefore natural to explore how star formation
depends on the galaxy rotational structure while maintaining {\it all}
the other properties of the model fixed. \footnote{Notice that changing
rotational support at fixed structure affects some aspect of the ISM
evolution, changing the amount of thermalization of the stellar
ejecta due to stellar velocity dispersion and to the relative motion
of stellar streaming rotational field and ISM velocity field. For an
extensive discussion see  \cite{2014MNRAS.439..823N} and \cite{2014MNRAS.445.1351N}; see also Kim \& Pellegrini (2012).} Accordingly, among the models presented in C22,
we focus our study on two different rotational realizations for each
of the LM, MM, and HM galaxies.  For each mass the two families are
named {\it constant-k} (and the models are indicated as HMc, MMc, and
LMc), and {\it exponential-k} (HMe, MMe, and LMe) for the following
reasons. As is well known, the azimuthal velocity field of a rotating
stellar system can be (and usually is) split in its ordered and
dispersion components by adopting a generalised Satoh (1980)
$k$-decomposition (see equation 11 in C22). In the constant-$k$
rotation case the Satoh parameter is constant over the galaxy body,
with $k=0$ corresponding to a non-rotating galaxy, with the flattening
totally supported by tangential velocity dispersion, while $k=1$
corresponds to a ``fast'' rotating galaxy (the isotropic rotator),
with the flattening fully supported by ordered rotation. In our HMc,
MMc, and LMc models we consider this latter possibility, i.e., they
are relatively fast spinning ETGs. In the exponential-$k$ rotation
case we considered instead the spatially-dependent Satoh parameter
\begin{equation}
  \ke(r)={\rm e}^{-r/\Rec},
\label{eq:satohkr}  
\end{equation}
so that the galaxy rotational support (and angular momentum injection
from the rotating stellar population) decreases significantly at large
radii, while in the central regions stars rotate almost as fast as an
isotropic rotator. Summarizing, in C22 and in the present study,
angular momentum injection in the ISM is maximal in constant $k>1$
models.

\subsection{Star formation}

Star formation is implemented in \texttt{MACER} as follows.  A first,
{\it necessary} condition to activate the star formation routines entails checking, grid by grid and at each time step, that the gas temperature is
  lower than $4\times 10^4$ K. Under this condition, $\drhost$ is
computed by considering the two channels mentioned in the
Introduction: the {\it Toomre instability channel}, and the {\it
  cooling/Jeans channel}. As we will see, in some circumstances the two
channels are both active; in this case $\drhost$ is defined as the
sum of the two rates (provided the total mass consumption rate in the
considered time step does not exceed a prescribed fraction of the
available mass in the considered grid).

The first channel of star formation is related to the Toomre
instability (e.g., see Toomre 1964, Binney \& Tremaine 2008). As is
well known, the local stability of a rotating, self-gravitating disk
can be understood in terms of the relative importance of temperature
(or velocity dispersion) vs. disk gravitational field (i.e. surface
density). Specifically, a self-gravitating gaseous disk is locally
stable when
\begin{equation}
    Q(R)\equiv {\cs (R) \kappa(R)\over \pi G \Sigmag (R)} > 1,
\end{equation}
where in our case $R$ is the distance from the galaxy center in the
disk plane, $\Sigmag$ is the gas surface density of the disk, $\cs$ is
the sound velocity, and $\kappa$ is the radial epicyclic frequency.
As an effect of disk instability, angular momentum of the rotating gas
is transferred outward due to a non-axisymmetric gravitational
torques; correspondingly, mass is transferred inwards and is eventually
accreted on the central SMBH\footnote{A semi-analytical algorithm
  mimicking this process is also included in \texttt{MACER}, and is
  essential for producing SMBH accretion in presence of ordered
  rotation.}.  Unstable over-densities in the infalling material (in
the present simulations, due to the imposed geometry, specifically
gaseous rings) trigger bursty phases of star formation, which then
decrease the surface density, increase $Q(R)$ above unity, and
re-stabilize the ISM, in a sort of self-regulation (e.g., see Bertin
\& Lodato 1999). In \cite{Gan_2019} the star formation rate
$\drhostT (R)$ associated with Toomre instability is captured in a
phenomenological way by
\begin{equation}
    \drhostT = \etaSFQ\Delta Q\rhog\Omega, \quad \quad \Delta Q = \max(1-Q, 0), \quad \quad \etaSFQ = 0.02,
\end{equation}
where $\Omega(R)$ is the disk angular velocity profile.

The second channel of star formation, activated only when the gas
density is higher than $10^5 \textrm{atom} \; \textrm{cm}^{-3}$,
compares the gas cooling timescale $\taurad$ and the dynamical
timescale $\taudyn$ of the ISM, and evaluates the associated star
formation rate $\drhostC$ defined as follows.
\begin{equation}
   \drhostC = \etaSFC {\rhog\over\tauSFC}, \quad \quad \tauSFC = \max(\taurad, \taudyn), \quad \quad \etaSFC = 0.01.
\end{equation}
The efficiency
parameters $\eta$ are consistent with results from previous simulations
\citep{Eve_2021} and observations \citep{Jiayi_2023}. The cooling timescale is computed as the ratio of total thermal energy
to the rate at which the gas cools, the latter of which is determined
by bremsstrahlung emission, Compton heating and cooling, and recombination
(Gan et al. 2019), with allowance for heating due to the central
flaring AGN.  The dynamical timescale is the minimum between the Jeans
collapse timescale $\taujeans = \sqrt{{3\pi}/{32G\rhog}}$ and the
rotational timescale $\taurot = 2\pi R/v_{rot}$. If the dynamical time
dominates, $\tauSFC \propto \rhog^{-1/2}$, whereas if the cooling
timescale dominates, $\tauSFC \propto \rhog^{-1}$, with the resulting
star formation rate proportional to either $\rhog^{3/2}$ or $\rhog^2$,
respectively. 

\section{Results}

\subsection{Gaseous and stellar disks from MACER simulations, and the
  global KS law}

A preliminary discussion of the global properties of the gaseous
rotating cold disks in ETGs (and of the associated stellar disks)
produced by \texttt{MACER} simulations can be found
in C22 (see Table 2 therein); here we consider several
additional properties of the circumnuclear disks, providing all the
information needed to address the focus of this paper. A complete
summary of the results is contained in Table 1 and Table 2. In the
Tables, the data are given for increasing galaxy mass, and for each
mass in order of increasing amount of ordered rotation. In particular,
in Table 1 we give the stellar and gaseous properties of the
circumnuclear disks, both at the end of the simulations (corresponding
to a simulation time of $\Delta t = 12$ Gyr, and to a final age of the
galaxy of $14$ Gyr), and time-averaged over $\Delta t$. In particular,
time-averaged quantities are indicated with a bar over the
corresponding symbol, so that the time-averaged values of the quantity
$f(t)$ is given by
\begin{equation}
\overline{f}\equiv{1\over\Delta t}\int_{t_0}^{t_0+\Delta t}f(t')\,dt', 
  \label{eq:timeav}
\end{equation}
where, as assumed in MACER, at the initial time of simulation the age
of the galaxy is $t_0=2$ Gyr. Therefore, for each of the six models,
the columns in Table 1 give the total mass $\Delta\Mstar$ of new stars
formed in the circumnuclear disk, the time-average $\Rdmsht$ of the
instantaneous half-mass radius of star formation $\Rdmsh(t)$, the
total mass $\Mdstar$ of the circumnuclear stellar disk at the end of
the simulation (where the difference $\Delta\Mstar-\Mdstar$ is
accounted for stellar evolution), and the final half-mass radius
$\Rdstar$ of the stellar disk. Then, for the properties of the
circumnuclear gaseous cold (defined as in C22 as gas with
$T\leq 5 \times 10^5$ K) disk, for each model we report the time-average
$\Mgt$ of the gas mass in the circumnuclear disk, together with the
time-averages $\Rght$ and $\Rgt$ of the gaseous disk half-mass and
truncation radii, respectively. The last two columns of Table 1
finally give the values at the end of simulation of the disk total
cold gas mass $\Mg$, and the associated truncation radius $\Rg$.

In Table 2 additional information about gas content and star formation
in the circumnuclear disks are reported to study the
{\it global} properties of the simulated KS relations. Accordingly,
for the six models we construct the equivalent surface star formation
rate, defined from the quantities in Table 1 as
\begin{equation}
<\dSigtot>\equiv {\Delta\Mstar\over 2 \pi\Rdmsht^2\Delta t}, 
  \label{eq:sigformav}
\end{equation}
i.e., the time-independent and spatially uniform star formation rate
of a fictitious starforming disk of half-mass radius $\Rdmsht$, forming a 
total stellar mass $\Delta\Mstar$ over the time $\Delta t$. We then
compute the time-average of the (spatially averaged) surface density of
the disk within the instantaneous half-mass radius, given by
\begin{equation}
\overline{<\Sigmag>}\equiv\overline{{\Mg(t)\over 2\pi\Rgh^2(t)}}. 
  \label{eq:siggasav}
\end{equation}
We finally evaluate the global, time-averaged star formation {\it
  efficiency}, naturally defined as
\begin{equation}
\epsilon_*\equiv \overline{{\dot\Mstar (t)\over\Mg (t)}}\simeq
{\Delta\Mstar\over\Mgt\Delta t}, 
  \label{eq:effstar}
\end{equation}
and, for comparison with the observed KS law, we also estimate the
global, time-averaged star formation {\it amplitude} as
\begin{equation}
<\Ampl>\equiv\log_{10}{<\dSigtot>} -1.4\log_{10}\overline{<\Sigmag>}, 
\label{eq:amplstar}
\end{equation}
where, following the common approach, $<\dSigtot>$ is in units of
$\Msun\kpc^{-2}\yr^{-1}$, and $\overline{<\Sigmag>}$ in
$\Msun\pc^{-2}$.

Before focusing on the properties of the resolved KS laws for the six
models, we list the main global results, beginning with Table 1. We immediately note a few robust trends. 1) The total mass of
newly formed stars in the circumnuclear disk $\Delta\Mstar$ is in the
range of $10^8 - 10^9\Msol$ and increases systematically with galaxy mass; at fixed galaxy mass, star formation is also larger for isotropic rotators
(subscript ``c'' in Column 1). Both these trends are quite
natural; in particular, the enhanced star formation in case of
substantial galaxy rotation confirms previous studies (Negri et
al. 2015, Gan et al. 2019) that angular momentum tends to
increase ISM cooling and successive instabilities. 2) The present day
mass $\Mdstar$ of the resulting stellar disks is obviously less than
$\Delta\Mstar$, a consequence of stellar evolution: notice that the
quite large difference between the two quantities (with a reduction
factor slightly larger than $\simeq 50\%$) in all the models is due to
the adopted top-heavy IMF in the simulations (see C22). 3) The present
day half-mass radius $\Rdstar$ of the stellar disk is found, for all
models, in the range $\simeq 100 - 300$ pc, being obviously larger for
the three isotropic rotator models: notice how well $\Rdstar$ matches
$\Rdmsht$ in Column 2, the time-averaged half-mass radius of star
formation in the disk.

The properties of the stellar disks reflect nicely the properties of
their gaseous component, the source of ISM material for star
formation: some systematic differences are however important and
should be noticed. 4) The present-day mass $\Mg$ of cold gas in the
disk, in each of the three different familiesof models, is
significantly larger (with values as high as $6\,10^{10}\Msol$) in the
isotropic rotators than in the corresponding low-rotation cases, where
in these latter cases $\Mg$ is only slightly larger than $\Mdstar$.
5) A similar comment applies to the size of the gaseous component of
the disks, as can be see by comparing Columns (4) and (7): in all
models, the stellar disks are obviously restricted to the central
regions of the parent gaseous disks, that can easily reach the kpc
scale.

All the previous results reflect in the global quantities concerning
star formation in the models, as reported in Table 2: in each of the
three families of galaxy models, the averaged surface density star
formation rate $<\dSigtot>$ is higher in the low-rotation models,
while the opposite happens for averaged cold gas surface density
$\overline{<\Sigmag>}$. The corresponding global, time-averaged star
formation efficiencies are in general of few percents: therefore, the
efficiency with which cold gas is converted to stars in the numerical
simulations is within the lower range predicted by
\cite{Kennicutt_Jr__1998} for spiral galaxies. The last, but
particularly important quantity, one of the focus points of the
present paper, is the amplitude of the simulated KS laws. From the
last column in Table 2, we found that $<\Ampl>$ is in the range
$(-5.1,-3.4)$; generally, these amplitudes lie lower than, but not extremely so, the Kroupa
IMF-corrected KS amplitude of measured for spirals/starbursts
\citep{Davis_2014}; lower amplitudes systematically correspond to models with low rotation of the stellar population. {\it Therefore, a
  reduction of the rotational support in ETGs at fixed galaxy mass,
  appears to decrease the amplitude of the resulting KS law, and one
  is tempted to conclude that specific angular momentum of the stellar
  population could be the leading parameter determining the
  normalization coefficient of the KS law}.


\renewcommand\arraystretch{1.4}
\begin{table*}
\centering 
\caption{Stellar and gas properties of the circumnuclear disks}
\vspace{2mm}
\begin{threeparttable}
\begin{tabularx}{\textwidth}{c@{\extracolsep{\fill}}ccccccccccc}
\toprule 
Model 
& $\Delta\Mstar$
& $\Rdmsht$
& $\Mdstar$
& $\Rdstar$
& $\Mgt$
& $\Rght$
& $\Rgt$
& $\Mg$
& $\Rg$\\
& $(10^8\Msun)$
& (\kpc) 
& $(10^8\Msun)$
& (\kpc) 
& $(10^8\Msun)$
& (\kpc) 
& (\kpc) 
& $(10^8\Msun)$
& (\kpc) \\
          & (1)      & (2)    & (3)   & (4)   & (5)      & (6)     & (7)    &  (8)     & (9)\\
\midrule 
LMe   & 4.8    & 0.1 & 2.1    & 0.1  & 7.3     & 0.2    & 0.7   & 2.4    &0.7    \\
LMc   & 6.8    & 0.2 & 3.0    & 0.3  & 72.4   & 2.8    & 7.7   & 60.3  &4.4     \\
MMe  & 12.2  & 0.1 & 5.3    & 0.1  & 14.0   & 0.3   & 1.0    &11.0   &0.5     \\
MMc  & 28.0  & 0.3 & 12.2  & 0.3  & 111.5 & 2.7   & 9.9    &46.6   &3.6      \\
HMe  & 28.9  & 0.1 & 12.5  & 0.1  & 27.2   & 0.5    & 1.85  &12.3  &0.6       \\
HMc  & 68.7  & 0.3 & 29.8  & 0.3  & 149.5  & 2.9   & 12.4  &57.8  &3.0      \\
\bottomrule 
\end{tabularx}
\begin{tablenotes}[para,flushleft]
  \footnotesize Star formation results for the 3 families of models in 
  order of increasing galaxy stellar mass, with radially dependent 
  (subscript ``e''), and isotropic (subscript ``c'') Satoh 
  decomposition for the streaming rotational velocity of the stellar 
  population; a bar over a quantity means time average over the 
  simulation time $\Delta t=12$ Gyr defined as in equation (\ref{eq:timeav}).  The 
  columns give: (1) total mass of new stars formed; (2) time-average 
  of the instantaneous half-mass radius of star formation $\Rdmsh(t)$; 
  (3) total mass of the circumnuclear stellar disk at the end of the 
  simulation: the difference $\Delta\Mstar-\Mdstar$ is accounted for 
  stellar evolution; (4) final half-mass radius of the stellar disk; 
  (5) time-average of the cold ($T\leq 5 \times 10^5$ K) gas mass in the 
  circumnuclear disk; (6) time-average of the half-mass radius of the 
  cold gaseous disk; (7) time-average of the truncation radius of the 
  cold gaseous disk; (8) final value of the cold gas mass in the 
  circumnuclear disk; (9) final value of the truncation radius of the 
  cold gaseous disk. 
\end{tablenotes}
    \end{threeparttable}
\label{results}
\end{table*}
\renewcommand\arraystretch{1.}
\vspace{5truemm}

\renewcommand\arraystretch{1.4}
\begin{table}
\centering 
\caption{Global parameters of the simulated KS in ETGs}
\vspace{2mm}
\begin{threeparttable}
\begin{tabularx}{\textwidth}{c@{\extracolsep{\fill}}ccccccccccccccc}
\toprule 
Model 
& $<\dSigtot>$
& $\overline{<\Sigmag>}$
& $\epsilon_*$
& $<\Ampl>$ \\
& $(\Msun\kpc^{-2}\yr^{-1})$
& $(\Msun\pc^{-2})$
& (Gyr$^{-1}$) \\
& (1) & (2) &(3) & (4)\\
\midrule 
LMe   & 0.5 & 2456.5 & 0.05 & -5.1\\
LMc   & 0.2 & 153.7 & 0.01 & -3.8\\
MMe  & 1.2 & 2580.3 & 0.07 & -4.7 \\
MMc  & 0.5 & 248.7 & 0.02 & -3.7 \\
HMe  & 3.6 & 1977.1 & 0.09 & -4.1  \\
HMc  & 1.4 & 332.5 & 0.04 & -3.4 \\
\bottomrule 
\end{tabularx}
\begin{tablenotes}[para,flushleft]
  \footnotesize Relevant parameters of the simulated KS law in the 
  simulated ETGs reported in Table 1. The columns give:
    (1) the equivalent (time-independent and spatially uniform) surface
  star formation rate $<\dSigtot>$ of a fictitious starforming disk of
  half-mass radius $\Rdmsht$, forming a total stellar mass
  $\Delta\Mstar$ over the time $\Delta t$, defined as in equation
  (\ref{eq:sigformav});
   (2) time-average of the average surface density of cold gas in the
  disk $\overline{<\Sigmag>}$, defined as in equation
  (\ref{eq:siggasav});
(3) time-average of the global star formation {\it efficiency}
$\epsilon_*$, defined as in equation (\ref{eq:effstar});
(4) {\it amplitude} of the KS star formation law, defined as in
equation (\ref{eq:amplstar}).
\end{tablenotes}
    \end{threeparttable}
\label{results}
\end{table}
\renewcommand\arraystretch{1.}
\vspace{5truemm}



\subsection{Resolved KS Laws}

Even though in principle star formation can happen everywhere over the
galaxy model, the simulations revealed that all star formation occurs
in rotationally supported, thin, cold disks in the galaxy equatorial
plane. To investigate the radial dependence of the star formation rate
$\dSigtot$ as a function of the surface gas density $\Sigmag$, i.e.,
the resolved KS law as produced by the \texttt{MACER} simulations, for
each of the six models we compute the mass of cold ($T\leq 5 \times 10^5$ K)
ISM, and the star formation rate, over radial annuli $(R,R+\Delta R)$,
and then we divide the results for $2\pi R \Delta R$. This computation
is repeated, for each annulus, with a sampling time of 1 Gyr over the
time $\Delta t =12$ Gyr spanned by the simulations, so that for each
of the analyzed models we construct a large set of points
$(\Sigmag, \dSigtot)$, corresponding to different radii in the disk,
and to different evolutionary times.

\begin{figure}
    \centering 
    \includegraphics[height = 170 mm, width = 170 mm]{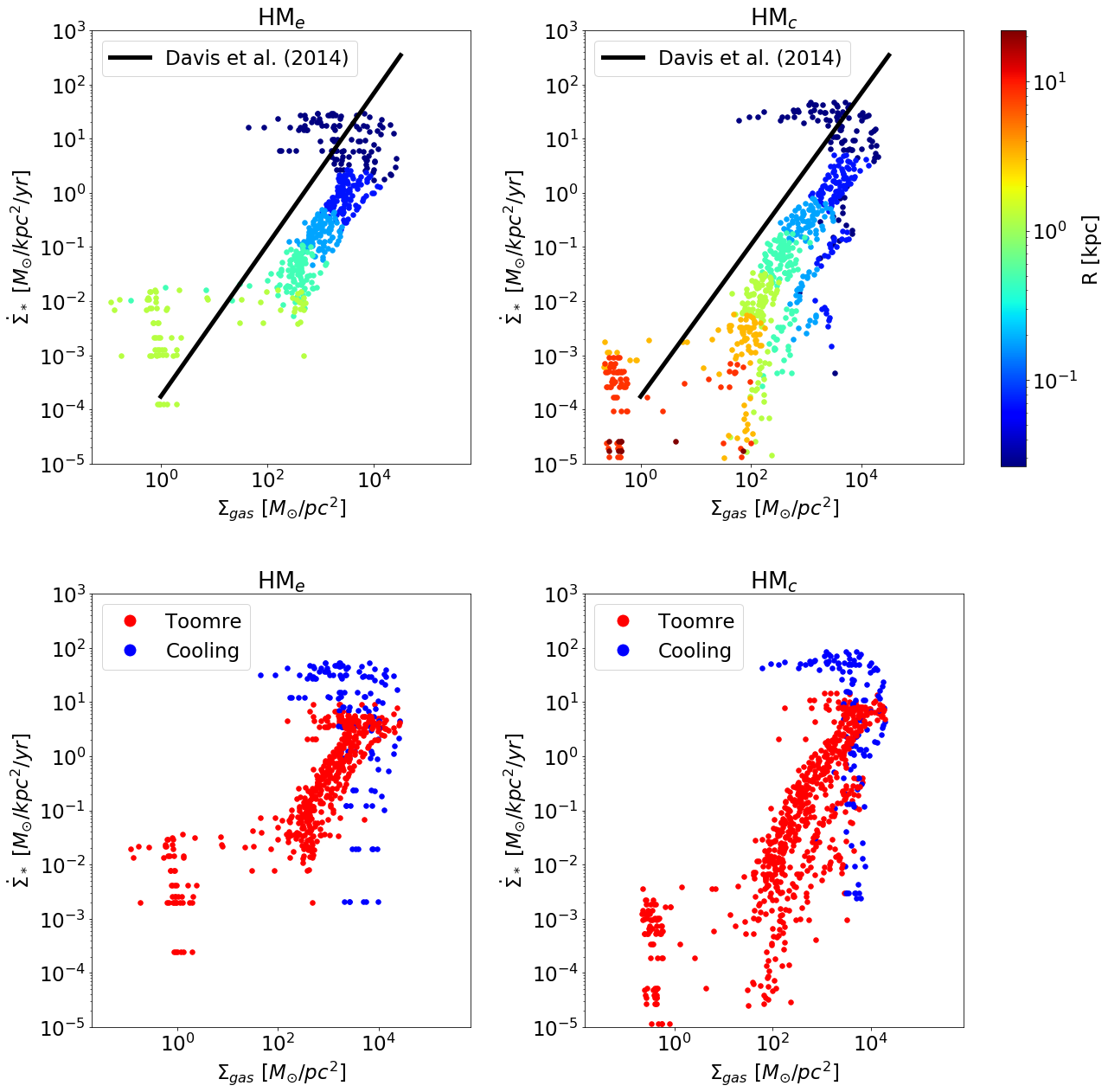} 
    \caption{Resolved star formation rate of the high-mass (HM) models for
    the low-rotation case with exponentially declining Satoh 
      $k(r)$ parameter with (HMe, left column), and for fast rotating 
      $k=1$ isotropic model (HMc, right column). In the top panels 
      local star formation rates $\dSigtot = \dSigQ + \dSigC$ (in 
      units of $\Msol$ kpc$^{-2}$ yr$^{-1}$) are given as a function 
      of the local disk gas surface density $\Sigmag$ (in units of 
      $\Msol$ pc$^{-2}$), with a sampling time 
      of 1 Gyr over the time span $\Delta t = 12$ Gyr, so the plots 
      represent also the history of star formation in the model.  For 
      reference, the solid line represents the \cite{Davis_2014}
      observed KS law in equation (\ref{eq:KSobs}), of slope $1.4$ and 
      amplitude $\Ampl=-3.76$.  In the bottom panels red and blue 
      points separate respectively the contribution of Toomre 
      instability ($\dSigQ$) and cooling/Jeans instabilities 
      ($\dSigC$) to the total star formation rate of each point in the 
      top panels.}
    \label{fig:HM}
\end{figure}

With the aid of Figure \ref{fig:HM} we begin by illustrating the
results for the two high-mass models HM, in the low rotation case with
the exponentially decreasing Satoh parameter $k(r)$ (HMe, left
panels), and for the fast rotating, isotropic rotator (HMc, right
panels) with constant $k=1$.  In the top panels, all the points
$(\Sigmag, \dSigtot)$ are plotted with a color scheme mapping the
distance from the galaxy center, with decreasing distance from red to
blue. A few trends are apparent. The {\it first} is just a
confirmation that the cold rotating disk extends more in the fast
rotating isotropic HMc model than in the less rotating HMe model, as
is obvious from the tail of red points in the top-right panel, absent
in the top-left panel. The {\it second} important result is that the
bulk of the points are nicely aligned parallel to the heavy solid line
showing the observed KS law
\begin{equation}
\dSigtot\simeq 10^{-3.76}\,\Sigmag^{1.4}, 
\label{eq:KSobs}
\end{equation}
as given by Davis et al. (2014), where the units are the same as in
Table 2, and the amplitude\footnote{Notice that the amplitude in
  equation (\ref{eq:KSobs}) is not the estimated time and space
  averaged amplitude $<\Ampl>$ in equation (\ref{eq:amplstar}). In particular, the relative sizes of amplitudes in equation (\ref{eq:amplstar}) for the low mass models are reversed from those in Figure 4, suggesting that cold gas is over counted for the exponential case - a rapid decrease in rotation leaves much of the gas unable to produce stars.}
$\Ampl=-3.76$, similar to what found by other authors. For example
\cite{Boquien_2011} report observed amplitudes in the range
$-3.83\lesssim \Ampl \lesssim -3.02$ for spiral galaxies when surface
densities/rates are computed in the same units as above.  The {\it
  third} result is that our resolved KS laws, albeit parallel to the
observed law, are clearly below it, with lower star formation rate at
given surface density of the cold gas in the disk, i.e. in the
simulations ETGs gaseous disks are predicted to be inefficient with
respect to observed KS law observed in disk galaxies by almost an
order of magnitude.  Notice that from Table 2 this is reflected by the
time-averaged, global value of $<\Ampl>\simeq -4.1$ for the
low-rotating HMe model, but not for the isotropic rotator HMc (where
$<\Ampl>\simeq -3.4$).  The {\it fourth} feature of the simulated
points to be noticed is the large scatter (in both models) of
$\Sigmag$ at approximately constant $\dSigtot$, both in the very
central regions, and at the edge of the gaseous rotating
disk. Physically, the scatter implies that at the very high (center)
and very low (disk outskirsts) star formation rates, the gas surface
density is only weakly correlated with star formation, with
significant differences in $\Sigmag$ at fixed $\dSigtot$ at the
center, and with almost uncorrelated $\Sigmag$ and $\dSigtot$ at large
radii. More quantitatively, star formation rate becomes abruptly
horizontal near the center, with
$\dSigtot\gtrsim 10^{9.5}\Msol\textrm{kpc}^{-2}\textrm{Gyr}^{-1}$.  At
larger radii ($\approx$ 2.5 kpc) we also see an abrupt threshold
cutoff at gas densities less than
$10^{-3} \; \textrm{g} \; \textrm{cm}^{-2}$. K89 remarks that a
similar behavior is a consistently observed phenomena for disk
galaxies, and that this cutoff threshold is typically found to occur
within the range of
$10^{-3} - 10^{-4} \; \textrm{g} \; \textrm{cm}^{-2}$, which is
incidentally very similar to what we observe for our ETG model in
Figure 1. That this feature is also present in our simulations is
encouraging regarding the veracity of our star formation recipe and
similarity of star-formation laws across different galaxy
types. Following K89, we argue that the universal cause for this
phenomena arises from stability considerations captured by the simple
Toomre star formation channel: at lower densities, the gas is stable
against large-scale perturbations, suppressing star formation. High
resolution observations of ETGs are essential, therefore, for
confirming or refusing the stunning similarity of star formation
dependence on gas density between disk and ETGs found in our work.

\begin{figure}
    \centering 
    \includegraphics[height = 170 mm, width = 170 mm]{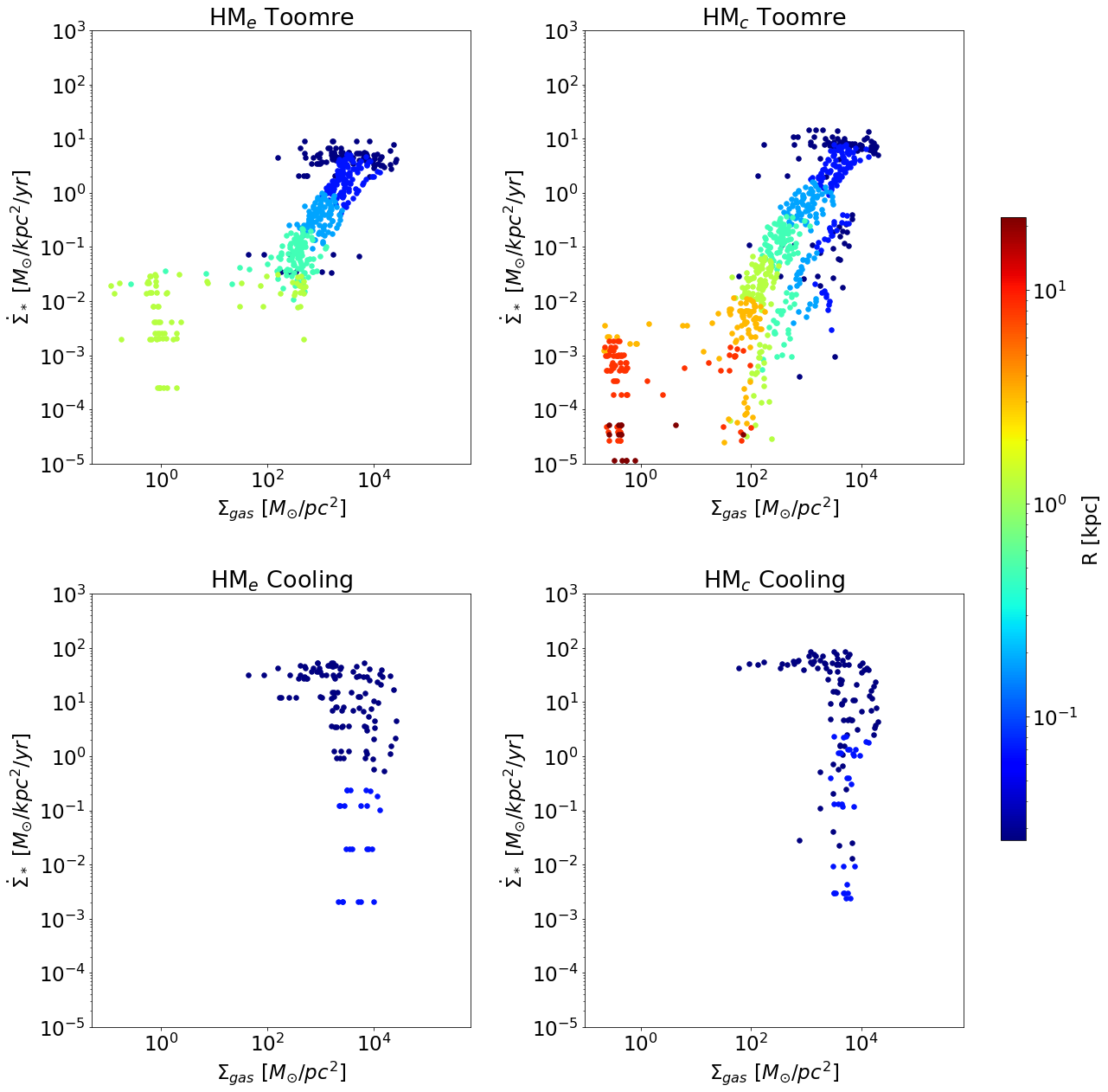}
    \caption{Star formation channels for models HMe (left) and HMc
      (right), color coded to show the position inside the rotating
      cold gaseous disk in the equatorial plane of the models. From
      bottom panels it is apparent how cooling/Jeans instability star
      formation is mainly confined to the very central regions, and
      producing the highest values of star formation. Toomre
      instabilities instead occur over all the disk.}
    \label{fig:HMchannels}
\end{figure}

All these trends are sufficiently interesting to merit a further
scrutiny. In fact, we recall that in the \texttt{MACER} treatment of
star formation, $\dSigtot = \dSigQ + \dSigC$, different possibilities
are considered, i.e., star formation can be produced by Toomre
instability, by cooling/Jeans instabilities, or by a combination of
the two, depending on the cold gas state (see Section 2). Therefore, a natural question is which, if either, of the two channels is dominant in
different regions of the $(\Sigmag,\dSigtot)$ planes. For this reason,
in the bottom panels of Figure \ref{fig:HM} we plot star formation resulting from different formation channels, indicating the star formation rate due to Toomre instability in red and the star formation rate due to cooling/Jeans instabilities in blue; all the points in the top panels corresponding to star formation with both channels active now
split in pairs of red and blue points. From the bottom panels, we
conclude that the bulk of the KS law obtained by
\texttt{MACER} is due to Toomre instabilities, which are
responsible for the large scatter of the points at the edge of the
cold disk. Conversely, star formation in the very central regions is
dominated by cooling/Jeans instabilities. Notice the almost vertical
strip of blue points at high values of $\Sigmag$: these points are not
missing in the top panels, they just represent a marginal star
formation well below the Toomre instability channel, and they
``disappear'' when added to the red points relative to the same
numerical cell and same sampling time. A complementary view of the
relative importance of Toomre and cooling/Jeans instabilities, where
the dependence of the main channels of star formation as a function of
the distance from the center can be clearly seen, is given in Figure
\ref{fig:HMchannels}, nicely confirming the conclusions above.

The dependence of the surface density profile on disk scale height, which rapidly
evolves over the disk, accounts for the reason for the different relevance of the two
channels in different locations of the gaseous disk. We identify the regime between
$2.5\times 10^{-1}$ and $2.5$ kpc as dominated by the Toomre channel,
which produces a SFR identical to the KS power law. K89, and more
recently \cite{Boissier_2003}, have both remarked that the Toomre
instability criterion remarkably reproduces SFR power laws in spiral
galaxies; despite the different gas properties and disk structure of
ETGs, the form of local star formation is the same in our
computations.

\begin{figure}
    \centering 
    \includegraphics[height = 170 mm, width = 170 mm]{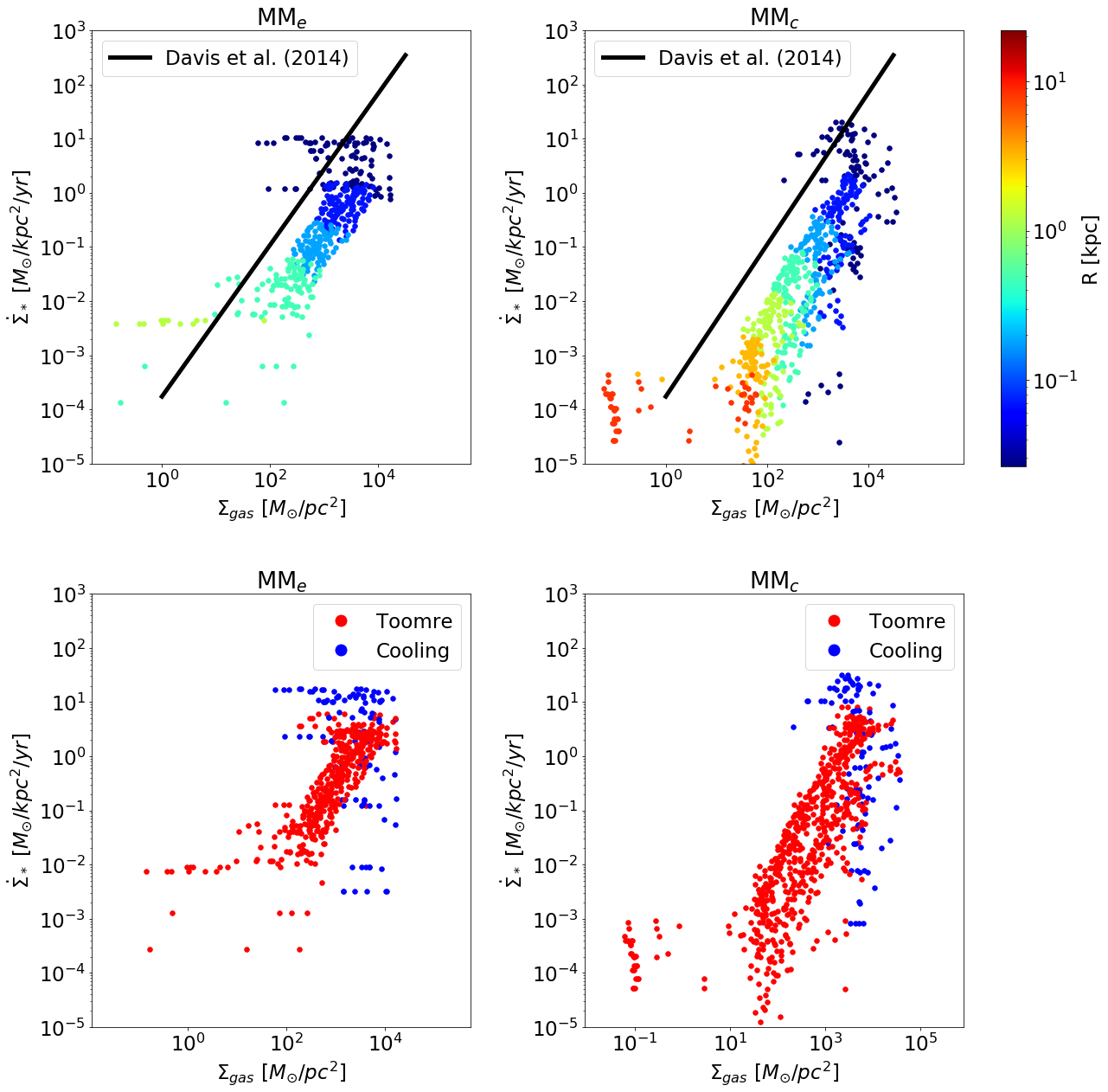}
    \caption{Resolved star formation rate of the medium-mass (MM) models, 
      for the low-rotation case with exponentially declining Satoh 
      $k(r)$ parameter with (MMe, left column), and for fast rotating 
      $k=1$ isotropic model (MMc, right column). In the top panels 
      local star formation rates $\dSigtot = \dSigQ + \dSigC$ (in 
      units of $\Msol$ kpc$^{-2}$ yr$^{-1}$) are given as a function 
      of the local disk gas surface density $\Sigmag$ (in units of 
      $\Msol$ pc$^{-2}$), with a sampling time 
      of 1 Gyr over the time span $\Delta t = 12$ Gyr, so the plots 
      represent also the history of star formation in the model.  For 
      reference, the solid line represents the \cite{Davis_2014}
      observed KS law in equation (\ref{eq:KSobs}), of slope $1.4$ and 
      amplitude $\Ampl=-3.76$.  In the bottom panels red and blue 
      points separate respectively the contribution of Toomre 
      instability ($\dSigQ$) and cooling/Jeans instabilities 
      ($\dSigC$) to the total star formation rate of each point in the 
      top panels.}
    \label{fig:MM}
\end{figure}

\begin{figure}
    \centering 
    \includegraphics[height = 170 mm, width = 170 mm]{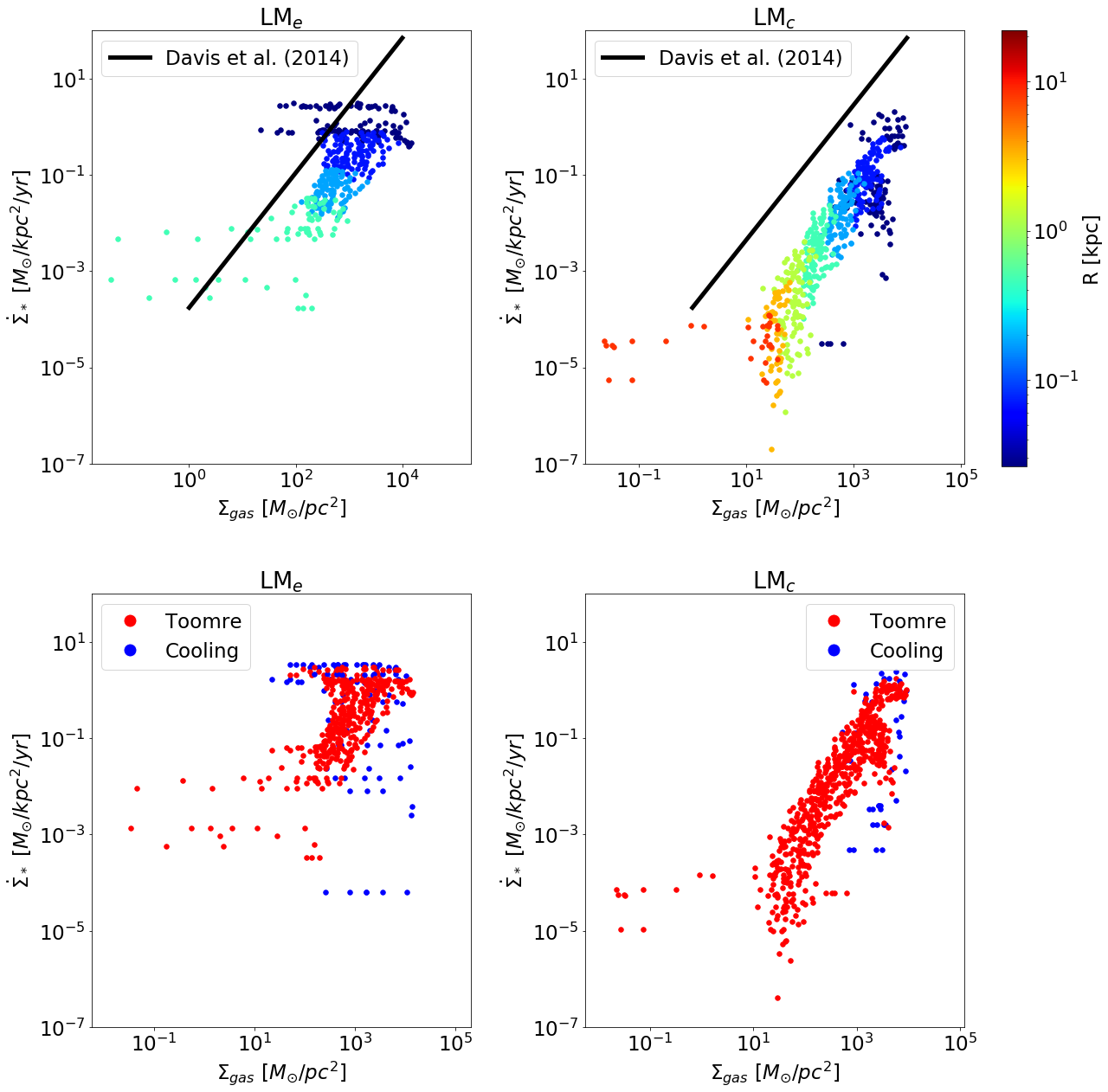}
    \caption{Resolved star formation rate of the low-mass (LM) models, 
      for the low-rotation case with exponentially declining Satoh 
      $k(r)$ parameter with (LMe, left column), and for fast rotating 
      $k=1$ isotropic model (LMc, right column). In the top panels 
      local star formation rates $\dSigtot = \dSigQ + \dSigC$ (in 
      units of $\Msol$ kpc$^{-2}$ yr$^{-1}$) are given as a function 
      of the local disk gas surface density $\Sigmag$ (in units of 
      $\Msol$ pc$^{-2}$), with a sampling time 
      of 1 Gyr over the time span $\Delta t = 12$ Gyr, so the plots 
      represent also the history of star formation in the model.  For 
      reference, the solid line represents the \cite{Davis_2014}
      observed KS law in equation (\ref{eq:KSobs}), of slope $1.4$ and 
      amplitude $\Ampl=-3.76$.  In the bottom panels red and blue 
      points separate respectively the contribution of Toomre 
      instability ($\dSigQ$) and cooling/Jeans instabilities 
      ($\dSigC$) to the total star formation rate of each point in the 
      top panels.}
    \label{fig:LM}
  \end{figure}

Figures \ref{fig:MM} and \ref{fig:LM} nicely confirm and extend the
global picture illustrated above also to different galaxy masses, with
simulated (resolved) star formation KS laws parallel to the observed
one, characterized by lower amplitude $\Ampl$, and by quite
significant scatter in the central regions and at the edge of the
circumnuclear gaseous disks.

\begin{figure}
    \centering 
    \includegraphics[height = 100 mm, width = 170
    mm]{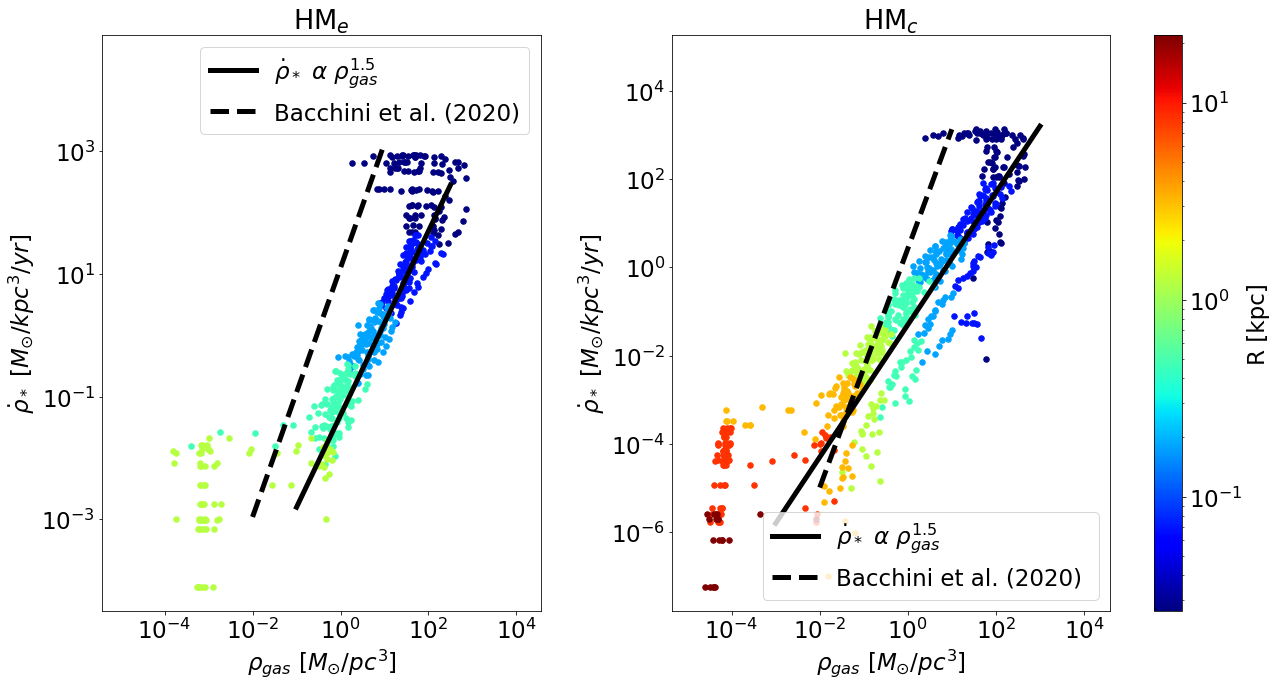}
    \caption{Volumetric star formation law for the low-rotation,
      high-mass model HMe (left), and the high-mass, fast rotating
      (isotropic) model HMc (right), a figure analogous to the top
      panels in Figure \ref{fig:HM}.  The total volumetric star
      formation rates $\drhost$ are given as functions of local gas
      density $\rhog$, with sampling time of 1 Gyr throughout the
      $\Delta t= 12$ Gyr time spanned by the simulation. The heavy
      solid line shows a {\it fit} to the $\drhost\propto \rhog^{1.5}$
      volumetric KS law, while the heavy dashed line is the {\it
        observed} relation derived for spirals and dwarf galaxies
      (equation 4.1 in Bacchini et al. 2020).}
  \label{fig:HMvol} 
\end{figure}

\subsection{The volumetric KS law}

As mentioned in \S1, {\it volumetric} KS laws have also been proposed
and investigated as alternatives to the standard KS law built from
surface densities. Owing to the thinness of the stellar disks and
difficulties of measuring volume densities in galaxies, surface
density star formation laws are more frequently used. However, surface
densities are frequently affected by projection effects and the
flaring of the disk thickness; thus, volumetric density relations may
offer more accurate insight into star formation laws while also being
more applicable for simulations. Due to these observational
difficulties, however, it is thought that conversion between surface
densities and volumetric densities is non-trivial (Bacchini et
al. 2019, 2020). Our numerical simulations allow us to study
volumetric power laws while avoiding observational 2D projection
effects.

We study the radial behavior of volumetric star formation laws by
averaging SFR and gas densities over radial annuli over 1 Gyr
timesteps throughout the $\Delta t =12$ Gyr of evolution. The two
  panels of Figure \ref{fig:HMvol} contain plots of the fit (heavy
solid line) of the obtained resolved volumetric KS law for the
  high mass models. We observe that the overall volumetric densities
follow a remarkable correspondence with the surface density
relationships in Figure \ref{fig:HM}, and that the fit is improved in
the low radii cooling regime with less horizontal spread. Using
Equation 4, a dynamical timescale on the order of $\rhog^{-1/2}$
explains the good $n \approx 1.5$ power law fit, consistent with some
observational work \citep{Jiayi_2023}: we conclude that in
\texttt{MACER} simulations it is the longer dynamical time rather than
the cooling time that dominates the star formation in this spatial
region, favoring $n \approx 1.5$ rather than the occasionally observed
$n \approx 2$  (Bacchini et al. 2020, heavy dashed lines). In any
  case, also the resolved volumetric KS laws obtained from the
  simulations confirm a lower efficiency of star formation at given
  gas density in the circumnuclear disks of the simulated ETGs. While
the volumetric KS relation corresponds with our physical prescription
at small radii, the surface density KS relation matches less well, as
noted in \S 3.2. This can be attributed to a rapidly flaring scale
height, as seen in the left hand panel of Figure \ref{fig:HMvol},
resulting in projection effects; we note, however, that Toomre
instability channel matches well in both cases. The threshold cutoff
observed at $\approx$ 2.5 kpc once again is attributed to Toomre
instability. Ultimately, we propose that surface-density SF laws,
which are easier to measure, can be translated into volumetric-density
SF laws, which describe the form of a local SF, given the close
correspondence between the two. More importantly, we once again assert
that the same empirical SF laws in disk galaxies can be applied
towards ETGs.  High-resolution observations of ETGs to confirm these
power law fits will thus be beneficial for applications towards
semi-analytical star formation recipes in simulations and for
understanding the local analytic form of star formation laws.

\section{Summary and Conclusions}

In this paper we present a detailed study of star formation laws in in
ETGs by analysing the results of high resolution numerical simulations
reported in C22 and performed with the latest version of our
\texttt{MACER} code.  For any axisymmetric ETG with some
level of ordered rotation in their stellar populations, both numerical
simulations and simple physical arguments have clearly established that the
ISM cools and develops large-scale instabilities, inevitably leading
to the formation of cold and rotating gaseous disks in the galaxy
equatorial plane. This is confirmed by recent observations
demonstrating the existence of massive gaseous disks in 50\% of ETGs
\citep{Young_2011}. Theoretically, such disks should be prone to star
formation, raising the question as to why their observed star
formation is so inefficient in comparison to disk galaxies
\citep{Negri_2014}.

Similar to what happens in rotating disk galaxy
models, our simulations lead to the formation of rotating, cold, gaseous
disks with kpc scale in the presence of galaxy rotation, gas cooling,
and angular momentum. We implement star formation over this cold gas
by considering two different channels: one based on Toomre
instability, and another determined by the longer of the local cooling
and dynamical (Jeans) times of the cold ISM. When the local values of
density and temperature of the ISM are respectively larger and smaller
than two prescribed threshold values, both channels are active, and
the star formation rate is their sum. Remarkably, our star formation
recipes result in a SFR-gas scaling law quite similar to the observed KS
relation.


The Toomre instability star formation channel reproduces an
$n \approx 1.4$ resolved KS star formation power law with a lower
cutoff threshold, which is analogous to what has been observed in disk
galaxies and can be attributed to gravitational stability
considerations. The cooling channel also obeys a power-law density
relationship, but deviates from the KS star formation law at higher
densities closer to the center of the galaxy, which may be attributed
to a rapidly flaring scale height. In preliminary 3D simulations more
recently performed, smaller scale, nonaxisymmetric fluctuations tend
to remove this small discrepancy. Volumetric star formation power laws
are also explored in effort to remove artificial effects of disk
flaring on surface density measurements, and we find a close
correspondence between the volumetric and surface density forms of the
KS power law. Globally, we find that our simulated ETGs lie
approximately parallel but with a lower normalization to
the same KS relation as disk galaxies


The similarity of the observed star formation power-laws in disk
galaxies with our simple numerical implementation in ETGs is
remarkable, implying that these KS laws provide simple recipes that
can be used in semi-analytical models of ETGs as well. The prediction
of the existence of small star-forming disks in ETGs encourages
high-resolution observations of galactic centers as well as further
work in understanding their dynamical properties and exploring the
form of local - rather than empirical - star formation laws to see if
our predictions correspond with reality.  Therefore, observational
checks of our predictions are imperative for evaluating these
conclusions and for determining the evolution of cool gas with respect
to the star formation inefficiency problem. ETGs are not quiescent in
our view and observations are need to check our SFR estimates: cooling
flows events that occur in rotating ETGs (from the non ejected gas)
should be observed as miniature versions of the standard observed disk
star forming regions in normal spiral systems, and the central regions
of the MW and M31.

\begin{acknowledgments} We thank Cecilia Bacchini, Eve Ostriker, 
    and an anonymous referee for important comments that improved the
    paper content and presentation.
ZG is grateful for the financial support from Princeton University via the subcontract to New Mexico Consortium.
\end{acknowledgments}

\nocite{*}
\bibliography{sample63}{}
\bibliographystyle{aasjournal}




\end{document}